\documentclass[10pt, a4paper]{article}

\usepackage{lrec-coling2024} % this is the new style

\usepackage{graphicx}

\usepackage{latexsym}

\usepackage{verbatim}
\usepackage{times}
\usepackage{latexsym}
\usepackage{graphicx}
\usepackage{amsmath}
\usepackage{booktabs}
\usepackage[T1]{fontenc}

\usepackage{multirow}
\usepackage{times}
\usepackage{latexsym}

\usepackage[T1]{fontenc}
\usepackage[utf8]{inputenc}

\usepackage{microtype}

\usepackage{inconsolata}

\title{Phishing Email Detection Using Inputs From Artificial Intelligence}

\name{P. Mithün, Genevieve Bartlett, Jelena Mirkovic, Marjorie Freedman} 

\address{Information Sciences Institute, Los Angeles,USA  \\
         mithun@isi.edu\\
         mithunpaul@arizona.edu
         \\
         }

\abstract{
Enterprise security is increasingly being threatened by social engineering attacks, such as phishing, which deceive employees into giving access to enterprise data. To protect both the users themselves and enterprise data, more and more organizations provide cyber security training that seeks to teach employees/customers to identify and report suspicious content. By its very nature, such training seeks to focus on signals that are likely to persist across a wide range of attacks. Further, it expects the user to apply the learnings from these training on e-mail messages that were not filtered by existing, automatic enterprise security (e.g., spam filters and commercial phishing detection software). However, relying on such training now shifts the detection of phishing from an automatic process to a human driven one which is fallible especially when a user errs due to distraction, forgetfulness, etc. In this work we explore treating this type of detection as a natural language processing task and modifying training pipelines accordingly. We present a dataset with annotated labels where these labels are created from the classes of signals that users are typically asked to identify in such training. We also present baseline classifier models trained on these classes of labels. With a comparative analysis of performance between human annotators and the models on these labels, we provide insights which can contribute to the improvement of the respective curricula for both machine and human training.
 \\ \newline \Keywords{phishing, phishing email detection training, machine learning} }

\begin{document}
\maketitleabstract
\section{Introduction}
\thispagestyle{empty} 
Social engineering attacks in the form of email communications that fool individuals into revealing personal or enterprise data are an on-going threat \citet{palmer,swig}.  While commercial solutions (e.g. Avanan, Barracuda Sentinel, FireEye etc.) exist for detecting attacks, even with such solutions attacks make their way into user inboxes. Thus, many organizations rely on training \citet{cofense,usc,harvard} to teach users to avoid dangerous behaviors (e.g. sharing passwords, clicking on links) and indications of a potentially malicious message (e.g. expressions of urgency, unfamiliar greetings, inconsistent email addresses). In such training, the user is asked to monitor for content that may occur regularly and in certain cases react differently. However, relying on such training shifts the detection of phishing from an automatic process to a hybrid one where the human part requires extra vigilance during mundane email checking. Further, relying on such training alone is fallible when a user errs due to distraction, forgetfulness, etc.

In this work we explore the degree to which approaches from NLP can be instead used to assist a user in performing the assessments such training would ask of the user. Specifically, we arrived at several labels that were derived from the types of signals that users are normally asked to look for during  anti-phishing training. Collectively, we refer to these labels as weak explainable phishing indicators (WEPI). The indicators are weak in that in isolation they do not necessarily indicate a malicious message. For example both urgency and link sharing can be part of a legitimate email. They are explainable because there is some rationale behind why they are used in phishing attacks, and thus they are less likely to be connected to a particular class of phishing attacks. 

To enable (and benchmark) such NLP-based approaches, we present an annotated corpus labeled for a selection of WEPIs. These WEPIs incorporate a mix of well-established labels and some novel labels arrived at from our data analysis. Specifically, we provide 32 WEPIs which we annotate in a corpus of 940 emails. Our corpus includes both malicious and legitimate emails since by their very nature, many of these indicators occur in both. Further, by analyzing the dataset we provide certain insights into the  frequency of occurrence of these indicators along with the venues where user training can be improved using them. We also train standard machine learning methods in identifying and classifying these WEPIs and by analyzing their performance we provide insights into what labels can be incorporated into an automated approach.

\begin{table*}[h!]
%\begin{tabular}{|l|l|l|r|}
\begin{tabular}{|p{15mm}|p{20mm}|p{97mm}|p{9mm}|}
\hline
\multicolumn{1}{|c|}{\textbf{Scope}} & \multicolumn{1}{c|}{\textbf{Short Name}} & \multicolumn{1}{c|}{\textbf{Description}} & \multicolumn{1}{c|}{\textbf{Count}} \\ \hline
\multirow{3}{*}{Message} & msgPa & Message from a Person Asking for performing an action & 594 \\ \cline{2-4} 
 & msgPo& \begin{tabular}[c]{@{}l@{}}Message in which a Person explicitly \\ represents an Organization as whole\end{tabular} & 195 \\ \cline{2-4} 
 & msgO & \begin{tabular}[c]{@{}l@{}}Message that is sent by an Organization \\ (not an individual)\end{tabular} & 184 \\ \hline
\multirow{16}{*}{Sentence} & sentA & Sentence mentions of an email Attachment & 36 \\ \cline{2-4} 
 & sentC & Sentence asks the recipient to Click on a link & 144 \\ \cline{2-4} 
 & sentI & Sentence contains an explicit Introduction by sender & 97 \\ \cline{2-4} 
 & sentM & Sentence talks about offering Money & 299 \\ \cline{2-4} 
 & sentPh & Sentence requests for scheduling a Phone call & 108 \\ \cline{2-4} 
 & sentPr & Sentence offers a product to be sold & 111 \\ \cline{2-4} 
 & sentR & Sentence talks about Recruitment (typically for employment) & 33 \\ \cline{2-4} 
 & sentSc & Sentence asks for  a  meeting or Scheduling information & 151 \\ \cline{2-4} 
 & sentSe & Sentence talks about offering a Service & 69 \\ \cline{2-4} 
 & sentUn & Sentence asks you to Unsubscribe (typically by clicking on a link) & 20 \\ \cline{2-4} 
 & sentO & Sentence requests the recipient to use a specific external product used by the sender's organization (e.g. zoom) & 36 \\ \cline{2-4} 
 & sentPa & Sentence asks or talks about PASsword & 26 \\ \cline{2-4} 
 & sentPo & Sentence has a Polite tone & 352 \\ \cline{2-4} 
 & sentUr & Sentence has an Urgent tone & 199 \\ \cline{2-4} 
 & sentUc & Sentence contains Uncommon URLs (as determined by an external list) & 138 \\ \cline{2-4} 
 & sentUnu & Sentence contains a URL that is Unrelated to the sender's or recipient's organization (if the domain occurs as part of the email metadata) & 59 \\ \hline
\multirow{10}{*}{Signature} & sig & Full signature block & 173 \\ \cline{2-4} 
 & sigA & Mailing address mentioned in the e-mail Signature & 22 \\ \cline{2-4} 
 & sigE & Email address mentioned in the e-mail Signature & 28 \\ \cline{2-4} 
 & sigF & Full name mentioned in the e-mail Signature & 133 \\ \cline{2-4} 
 & sigJ & Job title mentioned in the e-mail Signature & 32 \\ \cline{2-4} 
 & sigO & Represented organization mentioned in the e-mail Signature  & 89 \\ \cline{2-4} 
 & sigP & Phone number mentioned in the e-mail Signature & 52 \\ \cline{2-4} 
 & sigOf & Signoff used in the e-mail Signature & 89 \\ \cline{2-4} 
 & sigU & URL mentioned in the e-mail Signature & 9 \\ \cline{2-4} 
 & sigH & Social medial handle mentioned in the e-mail Signature & 0 \\ \hline
\multirow{3}{*}{\begin{tabular}[c]{@{}l@{}}Word or\\  Phrase\end{tabular}} & wordR & Mentions the recipient's affiliated organization & 0 \\ \cline{2-4} 
 & wordSl & Mention of sender's geographical location & 128 \\ \cline{2-4} 
 & wordSo & Mention sender's affiliated organization & 44 \\ \hline
 \end{tabular}
\label{label_frequency}
 \caption{ All the 32 WEPI labels, along with their corresponding linguistic scope, description and count in the annotated dataset.}
\end{table*}
\section{Related Work}

There have been several works that learn phishing email detection by incorporating cues from natural language processing to extract features in the e-mail and using statistical or neural network based.

For example, \citet{verma2012detecting} was one of the earliest such works where they present a comprehensive natural language based scheme to detect phishing emails using features that are invariant and fundamentally characterize phishing. Another common methodology is learning based filters that analyses a collection of labelled coaching
data (previously collected messages with upright evaluations) and understand the taxonomy of phishing emails as done by 
\citet{almomani2013survey,kim2011detecting,gupta2019machine}. \citet{aggarwal2014identification}
 exploits 
common features in phishing emails such like non-mentioning of the victim's name in the email, a mention of
monetary incentive and a sentence inducing the recipient to reply along with a header analysis methodology especially in phishing emails without links. \citet{yasin2016intelligent} in their work proposed a classification model using intelligent preprocessing phase for the extraction of various
features of email like email header, body, terms and frequency, by applying the techniques of data mining
and knowledge discovery for phishing emails or spoofed emails.

Several statistical methods have been employed also for identification of phishing email. For example \citet{harikrishnan2018machine} makes use of a distributional representation, namely TF-IDF for numeric representation of phishing mails along with presenting a comparative study of other classical machine learning techniques . In \citet{baykara2018detection} the authors developed an application to identify and detect the phishing element in text and message using the Bayesian classification algorithm with several databases. Some of the other prevalent statistics based phishing
email detection algorithms are supervised approaches,
such as support vector machines,logistic regression, Decision Trees and Naïve Bayes (\citet{verma2020email},\citet{kumar2020novel},\citet{niu2017phishing},\citet{hamisu2021detecting},\citet{junnarkar2021mail},\citet{swetha2019spam}, \citet{egozi2018phishing},\citet{gualberto2020answer},\citet{lee2020catbert},\citet{vinayakumar2019scalenet},\citet{janjua2020handling})

Another genre of methods that have been used is neural network based deep learning models. For example \citet{alhogail2021applying} propose a phishing email classifier model that applies
deep learning algorithms using a graph convolutional network (GCN) and natural language
processing over an email body text to improve phishing detection accuracy. \cite{abdullah2015research} uses simple feed forward network for phishing email detection.
Other deep learning methods used include using semantic graph neural networks, federated learning, RCNN models with multi level attention mechanisms, KNN, LSTM etc. 
\cite{prosun2022improved, manaswini2021phishing, lee2021d, hiransha2018deep, alotaibi2020mitigating, pan2022semantic, yaseen2021spam, thapa2023evaluation, halgavs2020catching, baccouche2020malicious, fang2019phishing, xiao2020malicious}

 From a pure natural language processing perspective our work is very similar to \citet{verma2012detecting}  and in-fact identifies a few more characteristics in the email that are invariant and fundamentally characterize phishing aiding in the ability of a model to distinguish between “actionable” and “informational” emails. However, our work is different because we are not proposing a phishing email detection algorithm, but instead show that there is an immediate need to modify the curricula of phishing email detection training, both for humans and machines. 

\section{Dataset Construction}
In this section we will describe various steps taken, and the rationale behind the creation of several of these labels and the subsequent dataset creation. All the 32 WEPI labels, along with their corresponding linguistic scope, description and count in the annotated dataset can be found in Table 1. 
 
\subsection{Label Selection}

To define a label set, we first reviewed several instances of anti-phishing guidance \citet{Ftc,Google, usc}, as well as known malicious phishing emails. Many of these guidelines urged users to be aware of certain common themes in phishing emails (e.g., urgency in the emails, or emails that requests the user to perform an unusual action or share some sensitive information), and some common contents that tried to confuse users by mimicking reputable external contents (e.g., the substitution of g00gle.com for google.com). 

\thispagestyle{empty} 
\subsubsection{Falsifiable and Verifiable Information}From the analysis of anti-phishing guidance and phishing e-mails we observed that the content of phishing e-mails can be classified into two broad classes: 
\begin{itemize}
    \item Content for which the topic or tone is explicitly tied to malicious emails. These included strong signals (e.g., requests for money) and weak signals (e.g., underlying urgent tones). 
    \item Content that can be falsified and/or recognized as a mismatch (e.g., email address in the signature). 
\end{itemize}

    \thispagestyle{empty} 
\textbf{Falsifiable Information:} One insight we gained from this analysis was that when mismatches appear, they are often a stronger signal of phishing than the content itself. For example, while a request for money is typically considered as a strong indicator of phishing, there are many instances where such requests can be legitimate. However, having an email address in your signature that is different than email address in the metadata is almost always a strong indicator of phishing. This is also an indicator of the difficulty faced by humans in confidently classifying an email as phishing or not. From this we arrived at the conclusion that the second category of falsifiable information can be further broken down into two sub-classes based on their verifiability as: 
    \begin{itemize}
        \item The alignment between the identity established in the text of an email and identifying features in the email metadata. (e.g., Does the  name of the \textit{organization} in the text of the email match the domain of the email address)?
    \item The alignment between the identity established in the text of an email and publicly available information about the sender. (e.g., Does the sender with the given name actually work in the organization that they claim in the signature?)
\end{itemize} 
  \textbf{Verifiable Information}: Further analysis of such verifiable information found in typical phishing and non-phishing e-mails made us arrive at the following insights :
\begin{enumerate}
\item Identifying the mismatch often relies on information that is difficult for a person to parse  easily (e.g. the email's metadata) or are information that is available externally (e.g., parent organization of the sender, which could be less available given historical emails). Hence, rather than annotate mismatches directly, we designed our e-mail annotation task to mark the components of a mismatch. Thus inspired by the warning that \textit{"A sender email address that does not match who the email claims to be from"}, we chose to annotate an email sender's affiliation claims (the label \texttt{wordSo} in Table 1) and the email address in a signature (\texttt{sigA}). Note that such an approach assumes the existence of a downstream verification algorithm (e.g. process the email metadata to determine if there is a mismatch).

\thispagestyle{empty} 
 
\item The signature block of an e-mail (\texttt{sig}), if it exists, is a rich source of verifiable information. Because of its formulaic (and white-space sensitive) structure, we treat it as a distinct scope. When present, we annotate the block as a whole and elements within it. Thus our annotation guidelines included asking annotators to identify physical mailing address (\texttt{sigA}), email address (\texttt{sigE}), full name (\texttt{sigF}), job title (\texttt{sigJ}), organization (\texttt{sigO}), phone number (\texttt{sigP}), the sign off (\texttt{sigOf}), urls (\texttt{sigU}), and social media handles  (\texttt{sigH}) if any were found in the Signature block of an e-mail.

\item When verifiable information appears in the body of the email, we are interested in access to the verifiable fact (e.g. the employer in an asserted employment relation between the sender and the recipient) and thus the scope of that annotation was decided to be word or phrase level. Thus, we annotate the affiliation relationships between an organization and the sender and recipient independently  (\texttt{wordR, wordSo}). Further, the location of  the sender (if mentioned in the text) is also annotated separately to capture another type of verifiable information (\texttt{wordSl}). Note that while some of these relations overlap with classic ACE-style relations, it may manifest themselves differently in email texts. For example, we expect to see the classic news-wire construction of \textit{organization title person name} (\textit{ACME researcher Jane Smith}) replaced with \textit{I'm Jane, a researcher working for ACME.}.
\end{enumerate}
\subsubsection{Content and Tone Information}
In designing the labels (and the subsequent annotation task) based on content and tone information, we found that identifying specific words was often overly prescriptive, but annotating the message as whole was overly broad. Thus we annotated several content and tone labels at the level of a sentence. These included aspects of tone (e.g., politeness \texttt{sentPo} and urgency \texttt{sentUr}) as well as intents of the sender (e.g., requests for money \texttt{sentM}, attempts by recruiters \texttt{sentR}) etc.

Finally, we annotate a handful of message level constructs that reflect the overall purpose of the emails. These are: Messages that appear to originate from an organization (\texttt{msgO}), Messages from an individual representing an organization (\texttt{msgPo}), and Messages from a person asking the recipient to perform an action (\texttt{msgPa}).

 Also as mentioned before, during annotation (and also during training models), these 32 indicators were further subdivided into four classes based on  linguistic scopes of annotation. The scope established what is marked (e.g., a few words, the whole sentence, the whole message, or the signature) and what unit of text needs to be fed as positive and negative examples to the model (specific details of which are  mentioned in the Appendix).

 Note that some of these labels are overlapping i.e the  same scope (e.g., sentence) can have multiple labels of the same kind. For example the same sentence might be asking the recipient to click on a link (\textit{sentC}) and also its URL can be pointing to an uncommon\footnote{ The URLs not found in the Alexa 1 million sites list \citet{alexa} was considered as uncommon.}URL (\textit{sentUc}).  On the other hand while requesting a phone call  can be considered a form of scheduling (\textit{sentSc}) only scheduling requests explicitly mentioning phone calls were marked as (\textit{sentPh}).

These are the rationale and intuitions based on which these 32 WEPI labels were arrived at.

\begin{figure*}[h]
 \includegraphics[width=1\linewidth]{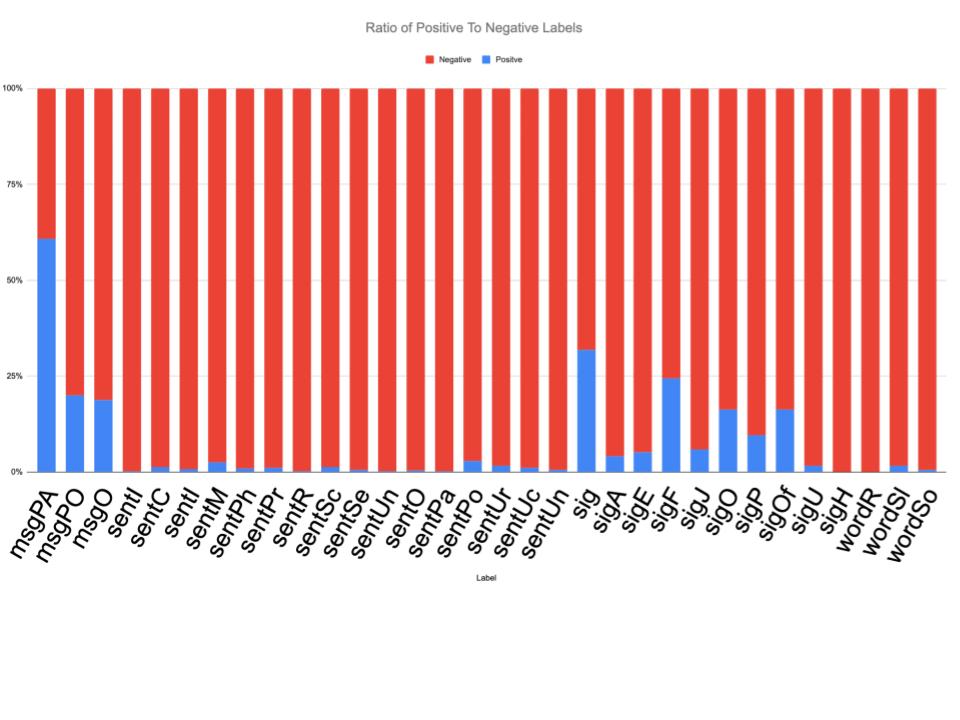}
    \caption{Ratio of positive to negative examples in the dataset for each label. The negative data points are higher than the positive data points, a design choice we made to faithfully reflect the distribution of labels across e-mails in a real-life inbox.}
  \label{ratio}
\end{figure*}

\subsection{Data Sources}

The emails for annotation were collected from 3 datasets: 336 emails from the Enron email corpus \citet{klimt2004enron}, 174 from the fraudulent email corpus of phishing emails \citet{fraud}, and another 430 emails taken from the inbox of our colleagues.

\subsection{Annotation Process}
 Annotation was done by a mix of students hired for the project and authors of the paper.  Annotators were paid above the minimum wage for the state of California. 
 
 The annotators were presented with an annotation interface that was created using Prodigy \citet{prodigy}. A set of pre-defined annotation guidelines, based on the intuition used in the label creation above,  were then presented to the annotators.
 
 The annotation process itself happened in training and testing phases. In the training phase, for each dataset we first we selected a small number of emails (e.g., 5) from each dataset and asked the annotators to annotate them. Then we calculated inter-annotator score of these annotations. If the score was below a certain threshold (e.g., 0.8 or above Cohen Kappa score for inter-annotator agreement between 2 annotators) we initiated a discussion between the annotators to arrive at a common understanding of the details of the methodology and thought process for assigning a given label. The annotation guidelines were also updated to reflect these improvements, if any.
 Once the inter-annotation agreement had improved on the training subset of the annotation dataset, we moved to the testing phase. In this phase, the annotators were presented with the e-mails that they had not seen before which were then annotated for the 32 WEPI labels. 

\section{Experiment Setup} 
To demonstrate the challenges presented by the dataset and analyze the potential future directions, we conduct a series of experiments which we describe in this section.

\label{sec:setup}

\begin{table*}[t]
\begin{center}
\footnotesize
\begin{tabular}{p{19mm}p{19mm}p{19mm}p{19mm}p{19mm}}
\toprule
 & {Bert-Base }&{Roberta-Base }&{Roberta-Large }&{Inter Annotator Agreement } \\
\midrule
sig&0.46&0.47&0.36&1\\
sigF&0.5&0.5&0.5&1\\
sigOf&0.67&0.14&0.19&1\\
sigO&0.5&0.17&0.22&1\\
sigP&0.5&0.23&0.23&1\\
sentM&0.08&0.17&0.05&0.96\\
sentPh&0.04&0.01&0.01&0.96\\
sentC&0.04&0.03&0.04&0.87\\
sentSc	&0.02&0.01&0.05&0.75\\
sentPo	&0.07&0.13&0.14&0.53\\
sentUr	&0.09&0.04&0.08&0.53\\
sentUc &0.02&0.02&0.03&0.98\\
sentPr &0.04&0.02&0.05&0.98\\
wordSl	&0.17&0.02&0.02&0.97\\
msgPa&0.84&0.74&0.77&1\\
msgPo	&0.61&0.25&0.34&0.98\\
msgO	&0.63&0.43&0.56&0.92\\
\bottomrule
\end{tabular}
\end{center}
\caption{ Performances (F1 score) of machine learning models and human annotators (Cohen Kappa inter-annotator agreement) on the labels in the dataset.  } 
\label{across_models}
\vspace*{-5mm}
\end{table*}

 \subsection{Training Data}

As shown in Table 1 the 32 WEPI labels, are classified based on linguistic scope. For all cases, when an extracted scope aligns with an annotated span, the passage is treated as a positive example for the aligned label.  Unaligned labels are treated as negative examples (Specific details of negative example creation process can be found in Appendix.) The distribution of positive to negative examples are shown in Figure \ref{ratio}. As illustrated in Figure \ref{ratio}, for most labels the dataset is highly unbalanced: the labels are rare and there are far more negative examples than positive. However, to ensure that the created dataset well reflects the real-world data distribution, we do not intervene in the label distribution and keep the unbalanced distribution as is. We treat 80\% of the data as train, 10\% as validation (for hyper-parameter tuning) and 10\% as test.  Also for the experiments presented here, we limit ourselves to those labels for which there were sufficient positive examples to support the fine-tuning of a model (refer Appendix for more details).

\subsection{ The Models}

Considering that pre-trained language models have dominated broad NLP tasks, we  use BERT \citet{devlin2018bert} and  RoBERTa \citet{liu2019roberta} as the backbone and build classification models on top of it, to provide simple but strong baselines. For each of the 4 linguistic scope based classes we set an additional classification head and we fine-tune the model for each of the classes. Further implementation details are given in Appendix. The corpus and the baseline models are available at our GitHub repository a link to which will be added after the blind review process.

\section{Results And Discussion }

Experimental results are shown in Table \ref{across_models} along with their corresponding inter-annotator agreement scores.

\subsection{Analysis}

 As shown in Table \ref{across_models}, the  subtle linguistic cues, captured by our labels, present a varying spectrum of difficulties, for both machine learning (ML) models and humans. Specifically, it can be seen that several emails where the sender is asking the recipient 
 for a thing or action (\texttt{msgPa}),  are easy for both machine and the human annotators to identify, as can be seen from the high F1 and high inter-annotator agreement scores. In these cases, a viable approach to phishing email identification would be to employ ML models to apply relevant labels to emails, and relieve human users of this cognitive burden. One could instead simply provide to human users some guidelines on how to interpret the attached labels.
 
 On the other hand there were several labels that the machine learning models struggled to identify, while humans did not (e.g., labels for which the scope was sentence). In such cases, user training could focus on asking users to identify this smaller set of labels, thus reducing the cognitive load. Furthermore, in the same scope, there were certain labels like the tone of urgency (\texttt{sentUr}) or politeness (\texttt{sentPo}) in a sentence which were equally difficult for both machines and humans to identify, as evident in lower F1 and lower inter-annotator agreement scores. In these cases, users may benefit from some targeted training, where they learn how to properly assign these labels, and how to estimate likelihood that such labels may indicate phishing. 
 \thispagestyle{empty} 
 
 Even comparing within ML models, we can see that certain labels are challenging for one model but easy for another.  For example, RoBERTa based models had difficulty understanding e-mails in which a person explicitly
represents an Organization as whole (\texttt{msgPo})  while BERT found it comparatively easier. This points to the need to employ ensemble learning for some labels, thus offloading these tasks from human users, while maintaining higher accuracies.

\subsection{Discussion}

Our work illustrates that  the  phishing email classification is a hard task from the perspective of natural language processing, encompassing the entire gamut of subtle difficulties natural language understanding entails, presenting difficulties for machines and humans alike.

While the goal of this exercise is not to create an automatic classifier for detection of phishing, but use the cues we provided in an anti-phishing training (e.g., providing the cues a user could be trained to look for). Note that in some cases this would require the context (e.g., of how the receiver knows the sender (or not) and if the sender is coming from an expected email). Ultimately that's knowledge that's hard to capture with an organizational-level "black-box" or other automatic approach. Thus we hope that our work will provide information on how to craft collaborative ways between machines and humans to identify suspicious emails that may contain phishing, while lowering human cognitive load. ML-derived labels can be presented to users along with their emails, and user focus could be directed only to assigning those labels that are challenging for ML-models to learn.

\thispagestyle{empty} 

Another possible utility of this corpus is explain-ability. For example, if a model trained on this corpus makes a decision that a given email is suspicious, we will be able to present explainable features to the end-user to substantiate this claim. (e.g., ``The reason this e-mail was tagged as suspicious the URL provided in the e-mail leads to not a well known website, possibly a phishing one.''). 

It can be argued that despite all user training, there is no guarantee that human end-user would always perform well against a new type of spear-phishing email presented with a novel nuance of urgency.  However, we hope to instill the relevance of such a label in the human learning experience. For example, even if these cues do not immediately succeed in a decision from the end user, we hope these will
act as a deterrent from taking an immediate debilitating action (which the sender of a phishing email typically wants), but instead taking a step back and raise an iota of doubt in the mind of the user.

On the other hand several of our cues may be useful in training machine learning models to augment the existing spam and phishing email filters. Specifically, we could possibly employ these models in a form of online learning to understand how often these cues are found in the emails received by the employees of a given organization and their respective significance. Minimally, an immediate application of our corpus can possibly be to serve as a fine-tuning venue for the large language models, which then might be able to be in turn deployed in developing models for the task of phishing email detection. Further, this corpus can possibly serve as a training dataset partially aiding in the creation of a downstream pipeline that can flag the emails that are missed by existing phishing filters. Also, when the model detects that the attack probability is sufficiently ambiguous, this model can possibly participate in the more expensive active detection by sending out specialized bots to update a message's attack likelihood through more in-depth analysis (of things like attachments, for example) or communication with the sender.

Another perspective on using our labels can be their permanence. Since they capture intent and tone of emails, and certain types of claims, rather than specific words used in emails, they should be resilient to the constantly changing strategies for phishing email creation.

\thispagestyle{empty}

\section{Conclusion}

In this work we investigated  use of higher-level, structured linguistic features of emails (rather than the actual words used in the email body)  for possible phishing email identification. To some extent the presence of such features is usually necessary for a phishing email's success. While these linguistic features could aid a downstream phishing classifier, our primary hypothesis is that some of these features may be challenging for machines to identify, while others may be challenging to humans. For this purpose, we annotated close to thousand emails for these labels, including phishing and benign  corporate emails, and present them as a dataset along with a trained model to serve as a baseline. Further, we show that these cues present a wide spectrum of difficulties for human and machine in the understanding of the natural language used in the phishing emails. These results offer hope that collaborative approaches between ML models and humans could be more accurate than separate identification, and could also lower human users' cognitive load when searching for phishing in emails. We hope our work will lead to the creation of models, with or without a human in the loop, that can minimally act a system that provides a warning to the recipient that will delay, if not prevent, the recipient from taking a debilitating action. Further, this can possibly lead to the creation of trained models that can help identify coordinated, large-scale social engineering attack campaigns and aid in attack attribution.

\section{Ethics and Limitation}
In the proposed work, we present an annotated dataset and trained models which identify certain indicators of phishing emails. We believe this study leads to intellectual merits that benefit from a reliable application of natural language understanding models in phishing e-mail detection and will further the training of phishing email identification curriculum. 2 out of the 3 datasets used were publicly available datasets. The 3rd dataset consisting of the emails was from the spam folder of the inbox of our colleagues with their permission.

Email in its raw form contains personally identifiable information. While de-identification could have been a solution, this can make the study of phishing in real contexts difficult. Here we address this ethical challenge by (1) using pre-existing released data; (2) in some cases using examples directly created for this effort; (3) for any data release anonymizing the data. Specifically, two of the datasets we use, (Enron dataset for benign emails and fraudulent email corpus for phishing emails) are pre-existing publicly available datasets. 

A second consideration is the labor involved in annotation.  Here, annotation was performed by a combination of the authors and students hired for the task.  Annotators were compensated for their time at a rate above the California minimum wage.

One limitation of our work is that the number of emails provided is possibly not enough to fine-tune a masked language model in an effort towards building a phishing classifier. However, this was a design choice we made due to the limitation of annotation resources. Also the negative data points are higher than the positive data points, in many cases affecting the ability of the ML model to identify them. This was a design choice we made to faithfully reflect the distribution of labels across e-mails in a real-life inbox.

\section{Acknowledgements}
This work was supported by the Defense Advanced
Research Projects Agency (DARPA) under the Active Social Engineering Defense (ASED) program. 
The authors would like to thank our student annotators Uma Durairaj, Neel Gupta and Zoe Zheng .
\section{Bibliographical References}\label{sec:reference}

\bibliographystyle{lrec-coling2024-natbib}
\bibliography{lrec-coling2024-example}

\end{document}